\documentclass[conference,a4paper]{IEEEtran}
\IEEEoverridecommandlockouts
\usepackage{cite}
\usepackage{amsmath,amssymb,amsfonts}
\usepackage{graphicx}
\usepackage{epstopdf}
\usepackage{textcomp}
\usepackage{balance}
\usepackage{subcaption}
\usepackage{cuted}
\usepackage{hyperref}
\usepackage{multirow}
\usepackage{amsmath}
\usepackage[english]{babel}
\usepackage{amsthm}
\usepackage{comment}
\usepackage{algorithm}
\usepackage{algpseudocode}

\usepackage[table]{xcolor}
\usepackage[left=1.65cm,right=1.65cm,top=1.93cm,bottom=4.3cm]{geometry} 
\makeatletter
\def\ps@IEEEtitlepagestyle{%
  \def\@oddfoot{\mycopyrightnotice}%
}
\def\mycopyrightnotice{%
  \begin{minipage}{\textwidth}
  \centering \scriptsize
  \copyright 2025 IEEE. Personal use of this material is permitted. Permission from IEEE must be obtained for all other uses, in any current or future media, including reprinting/republishing this material for advertising or promotional purposes, creating new collective works, for resale or redistribution to servers or lists, or reuse of any copyrighted component of this work in other works.
  \end{minipage}
}
\makeatother

\def\BibTeX{{\rm B\kern-.05em{\sc i\kern-.025em b}\kern-.08em
    T\kern-.1667em\lower.7ex\hbox{E}\kern-.125emX}}
\begin{document}

\title{Path Loss Modelling for UAV Communications in Urban Scenarios with Random Obstacles\\
}

\author{Abdul Saboor\textsuperscript{1},
        Zhuangzhuang Cui\textsuperscript{1},
        Evgenii Vinogradov\textsuperscript{1, 2},
        Sofie Pollin\textsuperscript{1}
\\\textsuperscript{1}WaveCoRE of the Department of Electrical Engineering (ESAT), KU Leuven, Leuven, Belgium
\\\textsuperscript{2}Autonomous Robotics Research Center, Technology Innovation Institute, Abu Dhabi, UAE
\\Email:\{abdul.saboor, zhuangzhuang.cui, sofie.pollin\}@kuleuven.be, evgenii.vinogradov@tii.ae}

\maketitle

\begin{abstract}
Path Loss ($PL$) is vital to evaluate the performance of Unmanned Aerial Vehicles (UAVs) as Aerial Base Stations (ABSs), particularly in urban environments with complex propagation due to various obstacles. Accurately modeling $PL$ requires a generalized Probability of Line-of-Sight ($P_{LoS}$) that can consider multiple obstructions. While the existing $P_{LoS}$ models mostly assume a simplified Manhattan grid with uniform building sizes and spacing, they overlook the real-world variability in building dimensions. Furthermore, such models do not consider other obstacles, such as trees and streetlights, which may also impact the performance, especially in millimeter-wave (mmWave) bands. This paper introduces a Manhattan Random Simulator (MRS) to estimate $P_{LoS}$ for UAV-based communications in urban areas by incorporating irregular building shapes, non-uniform spacing, and additional random obstacles to create a more realistic environment. Lastly, we present the $PL$ differences with and without obstacles for standard urban environments and derive the empirical $PL$ for these environments. 

\end{abstract}


\section{Introduction} 
Unmanned Aerial Vehicles (UAVs) are emerging as an enabler for the wireless industry due to their ability to provide on-demand connectivity \cite{polese2020experimental, gryech5systematic}. In recent years, UAVs functioning as Aerial Base Stations (ABSs) have demonstrated the ability to increase wireless coverage and network efficiency in smart cities. Compared to terrestrial Base Stations (BS), an ABS can adjust its altitude and coordinates to provide stable Air-to-ground (A2G) communication links \cite{saboor2021elevating}. Integrating higher frequencies in UAV-based ABSs, such as millimeter-wave (mmWave), further enables higher data rates for the users \cite{zhang2020optimized}. Furthermore, short mmWave wavelengths allow deploying large antenna arrays in compact spaces, making it suitable for UAVs \cite{xiao2020uav}.

However, short mmWave wavelengths make the channel propagation vulnerable to obstacles, such as buildings and trees. In such cases, the UAV channel power would mainly arrive at the Ground Users (GUs), including vehicles and Vulnerable Road Users (VRUs), by the Line of Sight (LoS) propagation \cite{khawaja2016uwb}. Durgin \emph{et al.} \cite{662630} reported that trees introduce 10 to 13 dB extra Path Loss ($PL$) at 5.85 GHz frequency, which is expected to be more at 28 GHz. Therefore, it is crucial to evaluate the Probability of LoS ($P_{LoS}$) and $PL$ for A2G communication links, primarily in urban environments with buildings and other random obstacles \cite{zhao2020efficient}. 

Various studies have tried to model A2G $P_{LoS}$ in urban environments \cite{ITU, hourani2014, holis2008elevation, Saboor2023plos, rev2}. Most of these models consider a Manhattan grid layout proposed by the International Telecommunication Union~\cite{ITU}. The Manhattan layout can be constructed using a tuple of built-up parameters, where all the generated buildings are square with equal widths. Furthermore, the space between buildings is identical. The key advantage of this approach is the layout simplification, making $P_{LoS}$ calculations tractable.

The Manhattan layout approach simplifies calculations but requires two important considerations. The first is real-world building diversity, where the area and space between all buildings are inconsistent, even in a Manhattan-style city. Furthermore, only a few buildings are square-shaped. The second is that these models overlook other obstacles in a city, like trees or streetlights, which can significantly impact mmWave signal propagation. 

Given the sensitivity of mmWave signals to foliage and other random obstacles in a city, developing a more comprehensive $P_{LoS}$ model for UAV-based A2G channels is important. Therefore, this paper presents a Manhattan Random Simulator (MRS) to examine $P_{LoS}$ in an urban environment constructed using ITU-defined built-up parameters. The MRS expands the simplified Manhattan grid model by incorporating irregular building shapes with non-uniform space between them. It also incorporates randomly placed obstacles such as trees and streetlights for $P_{LoS}$ estimation, making it more realistic for higher frequencies. The overall contributions to this paper are the following: 

\begin{itemize}
    \item We present MRS, which modifies the traditional Manhattan layout by incorporating random obstacles, irregular building areas, shapes, and space between them for estimating $P_{LoS}$.  
    \item We derived empirical $PL$ for three standard urban environments using simulated data. 
    \item We examine the impact of random obstacles, mainly streetlights and foliage, on $P_{LoS}$ and $PL$.
\end{itemize}

Section II of the paper overviews the traditional Manhattan layout and the proposed MRS-based layout. Section III explains how MRS works and derives empirical $PL$ for standard urban environments with random obstacles. Section IV shows the results, and Section V concludes the work.

\section{Urban Layouts using Built-up Parameters} 
The traditional simplified Manhattan layout uses a tuple of built-up parameters ($\alpha$, $\beta$, and $\gamma$) to generate a random city. The parameters $\alpha$ is the ratio of build area to total land area, $\beta$ is building quantity, and $\gamma$ is the Rayleigh scale parameter to generate random building heights according to Rayleigh distribution. Table \ref{tab1} provides the built-up parameters tuple for three standard urban environments. 

\begin{table}[!t]
\caption{ITU built-up parameters for urban environments.}
\centering
\begin{tabular}
{|l|c|c|c|}
\hline 
\textbf{Environment} & \textbf{$\alpha$} & \textbf{$\beta$ (buildings/km$^2$)}  & \textbf{$\gamma$ (m)} \\ \hline
Urban & 0.3 & 500 & 15 \\ \hline
Dense Urban & 0.5 & 300 & 20 \\ \hline
Urban High-rise & 0.5 & 300 & 50 \\ \hline
\end{tabular}
\label{tab1}
\vspace{-1em}
\end{table}

The main advantage of the traditional urban environment is its regularity and symmetry \cite{Saboor2023plos}, which makes it easy to simulate and model. All the buildings in this layout are squares with the same Width ($W$) \cite{ITU}. Furthermore, all the spaces between buildings are consistent, known as Streets ($S$), where a user can potentially reside. All buildings in such a layout follow symmetry, except buildings heights, which are random based on Rayleigh distribution.   

 However, the real cities following Manhattan's layouts are not symmetrical because of irregular building areas, heights, shapes, and spaces between them. Furthermore, such cities mostly have random obstacles like trees, poles, stations, and streetlights. This irregularity and presence of obstacles significantly affect $P_{LoS}$. Therefore, it is important to consider these parameters for modeling $P_{LoS}$.

 \subsection{Proposed Manhattan Random Layout}
\begin{figure}[!t]
\centerline{\includegraphics[width=.8\linewidth]{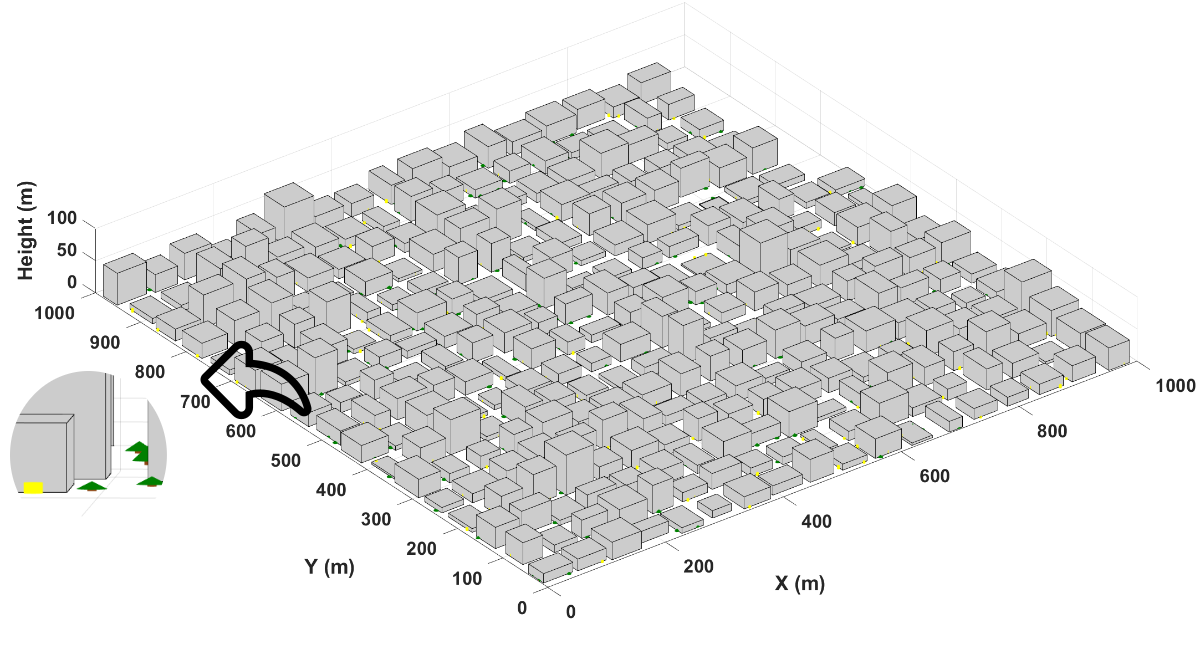}}
\caption{Proposed random Manhattan layout in MRS with varying building areas, shapes, and random obstacles.}
\label{ModifiedCity}
\vspace{-1em}
\end{figure}

The proposed random Manhattan layout divides the area into blocks. Unlike the traditional Manhattan layout with fixed building dimensions, it allows building width and length variations. Furthermore, it incorporates obstacles of trees and streetlights, as shown in Fig. \ref{ModifiedCity}. 

The total city area $A$ considered in MRS is 1 km$^2$, which is expandable to bigger areas. In the presence of $\beta$ buildings per km$^2$, the corresponding area occupied by buildings $A_b$ is calculated using $ A_b = \alpha \cdot A, \quad \alpha \in (0,1).$

The average area of each building $B_{avg}$ is then calculated as $B_{avg} = \frac{A_b}{\beta}$. We randomly fluctuate building widths and lengths around $B_{avg}$ to introduce building dimensions and shape variations. The proposed city considers square and rectangular-shaped buildings. Let $W$ represent the width and $L$ represent the length of a building. For square buildings, the dimensions are calculated using $W = L = \sqrt{B_{avg}}$. However, for a rectangle building with different $W$ and $L$, we introduce a shape factor as a random variable $\mathcal{R} \sim U(0.5, 1.5)$ to control variability in width. Thus, we have

\begin{equation}
W = \sqrt{B_{avg}} \cdot \mathcal{R}, \quad L = \frac{B_{avg}}{W}.
\end{equation}

The shape factor $\mathcal{R}$ ensures the building widths fluctuations at max/min 50\% of $B_{avg}$ to ensure realistic dimensions. Building heights are modeled using a Rayleigh distribution with parameter $\gamma$. The Probability Density Function (PDF) of building heights $h_b$ in an urban environment is given as\eqref{Bheights}:

\begin{equation}
\label{Bheights}
f(h) = \frac{h_b}{\gamma^2} e^{-\frac{h_b^2}{2\gamma^2}} \quad \text{for} \; h_b \geq 0,
\end{equation}
where the expected height and variance of building heights can be estimated using the following equation, 

\begin{equation}
\mathbb{E}[h_b] = \gamma \sqrt{\frac{\pi}{2}}, \quad \mathbb{V}(h_b) = \frac{4 - \pi}{2} \cdot \gamma^2.
\end{equation}


Trees in MRS are modeled as a combination of conical and cylindrical obstacles with radius $r_{T}$ and height $h_T$. The $h_T$ is uniformly distributed between $U(2, 5)$~m while $r_T \sim U(0.5, 1.5)$~m. The reason for keeping relatively smaller $h_T$ and $r_T$ is that the trees around the streets in urban environments are generally small. The trunk of a tree is modeled as a cylinder that takes 20\% of $h_T$ ($h_{Trunk} = 0.2 \times h_T$) and 10\% of $r_T$ ($r_{Trunk} = 0.1 \times r_T$). The remaining 80\% height represents foliage in the form of a cone with $r_T$ at the bottom, as visualized in Fig. \ref{ModifiedCity}. In contrast, the street lights are cylindrical obstructions with height $h_S  \sim U(2, 5)$~m and fixed radius $r_S = 0.1$~m.       

In the proposed Manhattan Random layout, trees and streetlights are distributed randomly in the city, around the edges of buildings at a distance $d_o$, ensuring that all these obstacles are placed on the sidewalk to represent a realistic scenario. Lastly, $n_{GU}$ ground users are randomly placed within the city's open space. The coordinates of each user $(x_u, y_u)$ are drawn uniformly from the area, subject to the condition that the user does not collide with any obstacles in the city, expressed as $(x_u, y_u) \notin \{ \text{Buildings}, \text{Trees}, \text{Streetlights} \}$

\section{$P_{LoS}$ Estimation using MRS}
Algorithm \ref{algo1} explains $P_{LoS}$ calculation process, where the MRS starts by generating a Manhattan layout consisting of varying building areas, shapes, heights, and spaces between them using built-up parameters in Table \ref{tab1}. Later, it randomly places $n_{trees}$ and $n_{lights}$ at $d_o$ from a random building side. The value of $d_o$ is 1.5~m to ensure the obstacles reside around the sidewalks. The third step involves placing random $n_{GU}$ within the city by avoiding obstacles.

After that, the MRS randomly generates an ABS with coordinates $(x_{ABS}, y_{ABS}, h_{ABS})$. Since a particular GU and ABS have fixed $(x, y)$ coordinates, the distance between them remains constant. However, $h_{ABS}$ is dynamic depending on elevation angle $\theta$, where $\theta$ ranges from $0^\circ$ to $90^\circ$. Here, $\theta$ is defined from plane XY of GUs to the Z axis of ABS. The next step is to check LoS availability by determining if any building, tree, or streetlight intersects the line connecting ABS and the ground user. If there are $n$ obstructions between ABS and GUs, the MRS compares the blockage height $h_{Blockage}(i)$ of each $ith$ building with corresponding obstacle height $h_{obstacle(i)}$ using\cite{simppaper}: 
\begin{equation}
\label{obstruction}
    h_{Blockage}(i) = h_{ABS} - \frac{r_{i}\times (h_{ABS}-h_{GU})}{r},
\end{equation}
where $r$ is the distance between ABS and the user, while $r_{i}$ is the distance between ABS and $ith$ obstacle. $h_{GU}$ is the ground user's height set to 1.5~m to represent both pedestrians and vehicles in the city. The MRS compares each $h_{obstacle(i)}$ with $h_{Blockage}(i)$, and the link is only considered LoS if all $h_{Blockage}(i) > h_{obstacle(i)}$. Otherwise, it is Non-LoS (NLoS). In the end, average $P_{LoS}$ is computed by dividing the sum of LoS values by the number of observations for each elevation angle.

\begin{algorithm}[!t]
\caption{$P_{LoS}$ Calculation for ABS and GUs.}
\small
\label{algo1}
\textbf{Input:} $\alpha, \beta, \gamma$, $n_{GU}$, $n_{cities}, n_{trees}, n_{lights}, d_o$ \\
\textbf{Output:} Average $P_{LoS}$ against all elevation angles
\begin{algorithmic}[1]   
    \For{each city $i = 1$ to $n_{cities}$}
        \State Generate Manhattan Random city using $\alpha, \beta, \gamma$
        \State Place $n_{trees}$ and $n_{lights}$ at $d_o$.
        \State Randomly place $n_{GU}$ and an ABS within the city by avoiding obstacles
    \EndFor
    
    \For{each user $j = 1$ to $n_{GU}$}
        \For{each elevation angle $\theta$ from $0^\circ$ to $90^\circ$}
            \State Calculate $h_{ABS}$ and $r$ for each $\theta$ against a GU
            \State Check LoS by verifying if any building, tree, or streetlight intersects the ABS-GU line using the equation \eqref{obstruction}
            \If{LoS exists}
                \State Increment $los(\theta^\circ)$
            \EndIf
        \State Increment $count(\theta^\circ)$
        \EndFor
    \EndFor
    
    \For{each angle $\theta$}
        \State Calculate $P_{LoS}(\theta^\circ)$: \quad $P_{LoS} = \frac{los(\theta^\circ)}{count(\theta^\circ)}$
    \EndFor
    
    \State \textbf{Return} $P_{LoS}$ \Comment{Average $P_{LoS}$ for each elevation angle}
\end{algorithmic}
\end{algorithm}

\begin{figure*}[!t]
  \centering
    \begin{subfigure}[b]{0.30\linewidth}
    \includegraphics[width=\linewidth]{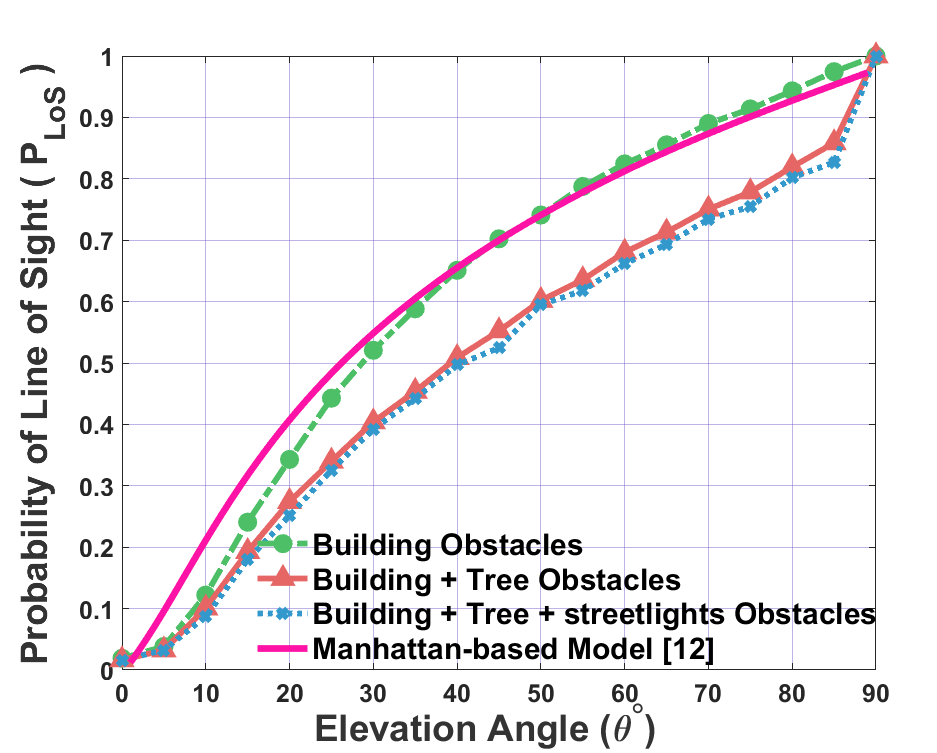}
    \caption{Urban Environment}
    \label{Urban}
  \end{subfigure}
  \centering
  \begin{subfigure}[b]{0.30\linewidth}
    \includegraphics[width=\linewidth]{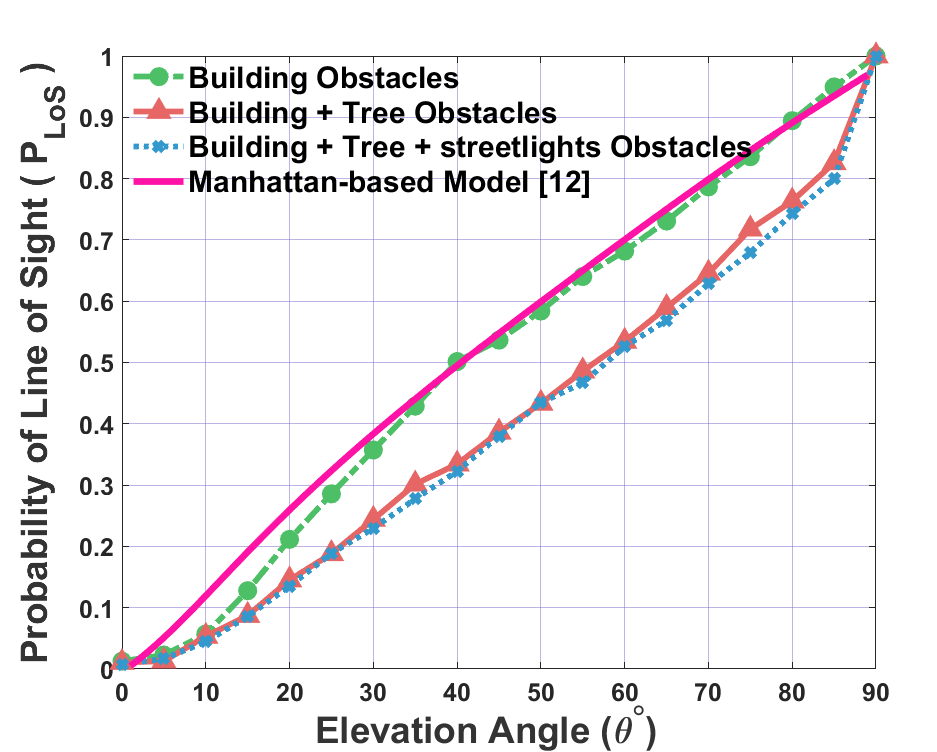}
    \caption{Dense urban Environment}
    \label{DUrban}
  \end{subfigure}
  \centering
  \begin{subfigure}[b]{0.30\linewidth}
    \includegraphics[width=\linewidth]{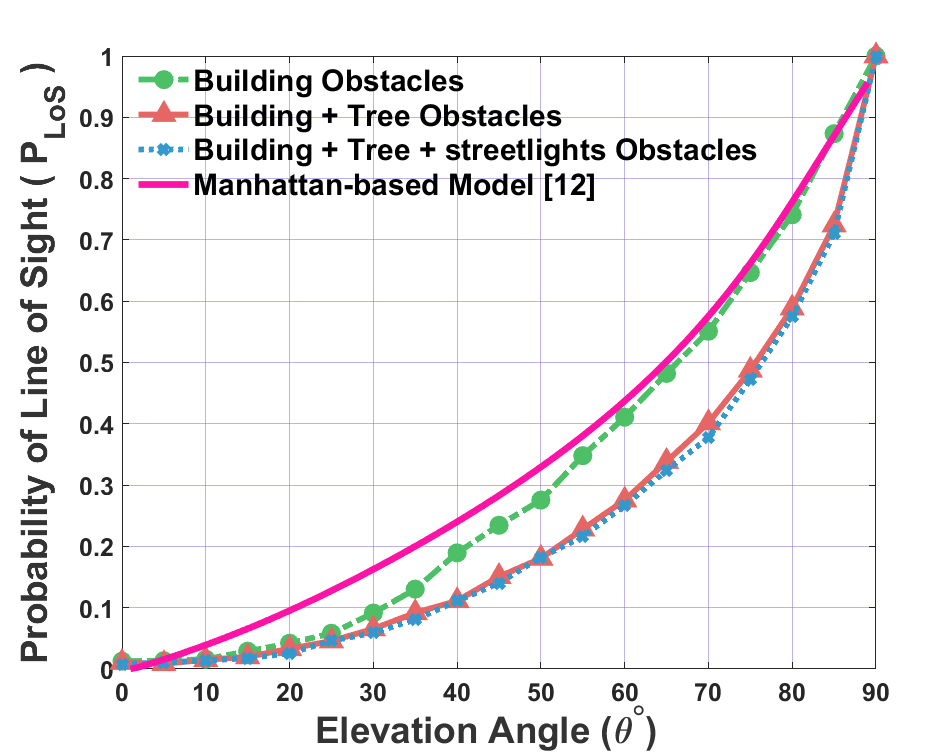}
    \caption{High-rise Environment}
    \label{HRise}
  \end{subfigure}
  \caption{$P_{LoS}$ in standard urban environments with multiple obstacles, where $n_{trees} = 200$, and $n_{lights} = 500$. }
  \label{PLoS curves}
  \vspace{-1em}
\end{figure*}

\vspace{-.5em}

\subsection{Empirical Path Loss Model}
The MRS provides $P_{LoS}$ against elevation angle ($\theta$) in the presence of multiple obstructions for any tuple of built-up parameters. For $N$ types of obstructions between ABS and GUs, the total $PL$ can be calculated using:
\begin{equation}
\label{GPL}
  PL = P_{LoS} \times PL_{LoS} + \sum_{i=1}^{N} P_{NLoS, i} \times PL_{NLoS, i}
\end{equation}

\noindent
where $PL_{LoS}$ represents $PL$ in the presence of LoS link. In contrast, $P_{NLoS, i}$ and $PL_{NLoS, i}$ are NLoS probability and $PL$ of $ith$ obstruction. In the presence of three types of obstructions: buildings, trees, and streetlights, the equation \eqref{GPL} can be rewritten as:
\begin{equation}
\label{PL}
\begin{split}
    PL = P_{LoS} \times PL_{LoS} 
    + P_{NLoS, B} \times PL_{NLoS, B} \\
    + P_{NLoS, T} \times PL_{NLoS, T} 
    + P_{NLoS, S} \times PL_{NLoS, S},
\end{split}
\end{equation}
where $P_{NLoS, B}$, $P_{NLoS, T}$, and $P_{NLoS, S}$ are the NLoS probabilities due to buildings, trees, and streetlights, respectively. Similarly, $PL_{NLoS, B}$, $PL_{NLoS, T}$, and $PL_{NLoS, S}$ are corresponding $PL$ due to building, tree, and streetlight obstructions, respectively. $P_{NLoS, B} + P_{NLoS, T} + P_{NLoS, S} = 1 - P_{LoS}$ is the total NLoS probability. The $PL_{LoS}$ is mainly modeled as Free Space Path Loss (FSPL), while $PL_{NLoS, B}$ can be estimated by the model proposed in \cite{6834753} at 28 GHz:
\begin{equation}
 \begin{split}
     PL_{LoS} = FSPL = 61.4 + 20\log_{10}(d) \quad \text{[dB]},   \\
     PL_{NLoS, B} =  72 + 29.2\log_{10}(d) \quad \text{[dB]}.
\end{split}   
\end{equation}

where $d$ is the 3D distance between ABS and GU. The $PL_{NLoS, T}$ is estimated using the following equation:
\begin{equation}
    PL_{NLoS, T} = FSPL + T_{Att}
\end{equation}
where $T_{Att}$ is additional attenuation caused by the trees\cite{ITUvege}, which can be computed using the equation \eqref{Att}. 

\begin{equation}
\label{Att}
   T_{Att} = R_{\infty}d_t + k \left( 1 - \exp \left( \frac{-(R_0 - R_{\infty})}{k} d_t \right) \right)
\end{equation}
where $R_0 = af$ and $R_{\infty} = b/f^c$ are the initial and final slopes. The attenuation factor $k$ is then calculated as:
\begin{equation}
\label{kfactor}
\small
k = k_0 - 10 \log_{10} \left( A_0 \left( 1 - e^{\left( -\frac{A_{min}}{A_0} \right)} \right) \left( 1 - e^{\left( -R_f f \right)} \right) \right)
\end{equation}
where $a = 0.2, \quad b = 1.27, \quad c = 0.63, \quad k_0 = 6.57, \quad R_f = 0.0002, \quad A_0 = 10$ are empirically defined for the leaf scenario \cite{ITUvege}. For A2G channels, leaves are more likely to obstruct the link between ABS and GUs than stems. Therefore, we consider attenuation by leaves only. $A_{min}$ is the minimum illumination area, which, in our case, mainly depends on the Fresnel zone. The Fresnel zone radius $r_{F}$ can be calculated using the formula:

\begin{equation}
r_{F} = \sqrt{\frac{\lambda \cdot d_1 \cdot d_2}{d_1 + d_2}}
\end{equation}

Where $\lambda$ is the wavelength, $d_1$ and $d_2$ are the distances from the transmitter and receiver to the point of interest. The minimum illumination area $A_{min}$ is then:
\begin{equation}
A_{min} = \min(2r_{F}, 2r_T) \times \min(2r_{F}, 2r_T).
\end{equation}

Our findings reveal that $P_{NLoS, S}$ is very low in the presence of 500 streetlights, making the impact of $PL_{NLoS, S}$ negligible. Therefore, we exclude this impact to maintain the simplicity of the model. The empirical $PL$ model for three standard urban environments is given in a general $A-B$ path loss model in \eqref{empirical}, 
\begin{equation}
\label{empirical}
     PL = A + 10B\log_{10}(d) \quad \text{[dB]}   \\
\end{equation}
where $A$ and $B$ for these urban environments are given in the following sections.

\section{Simulation Results and Analysis} 
This section compares the $P_{LoS}$ and $PL$ in the presence of multiple random obstacles in standard urban environments. Each simulation deploys an ABS and 100 GUs in a city, each yielding 90 points representing $P_{LoS}$ for each elevation angle. This simulation is repeated with 30 cities; thus, 270K points between ABS-GU are taken for averaging.    

\begin{figure}[!t]
    \centering
    \begin{minipage}{0.46\linewidth}
        \centering
        \includegraphics[width=\linewidth]{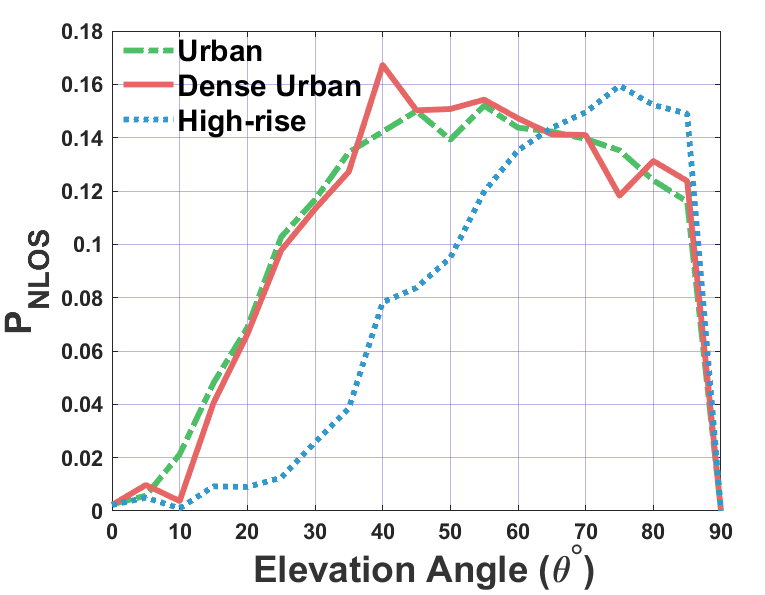}
        \caption{Additional $P_{NLoS, T}$ for 200 trees for three urban environments.}
        \label{Diff}
    \end{minipage}
    \hfill
   \begin{minipage}{0.46\linewidth}
        \centering
        \includegraphics[width=\linewidth]{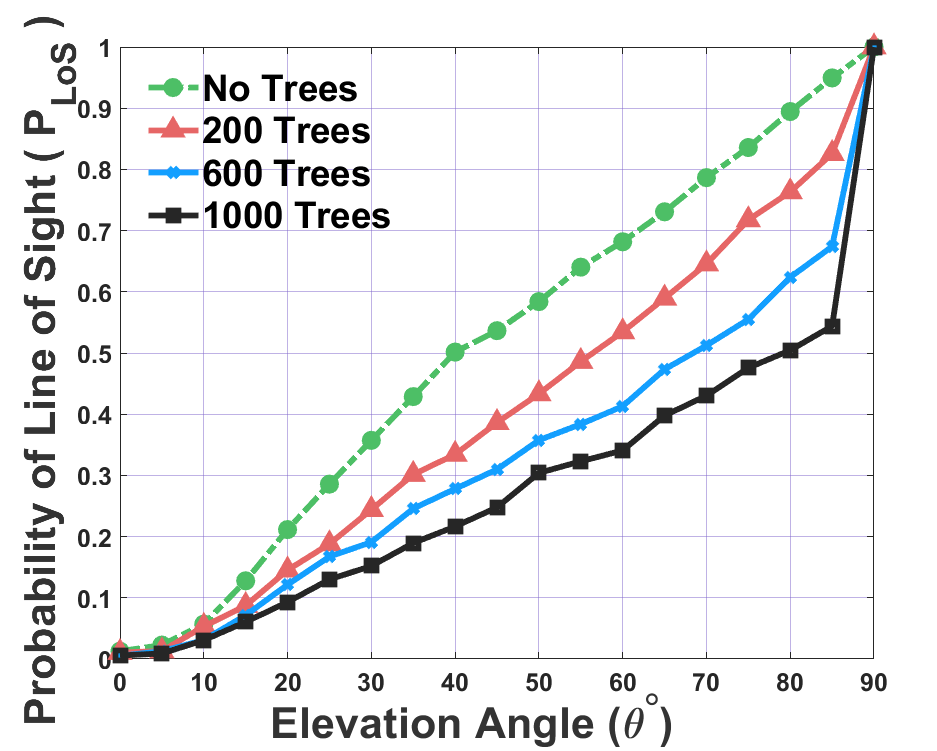}
         \caption{Impact of tree density on $P_{LoS}$ in dense urban environment. }
        \label{Density}
    \end{minipage}
    \vspace{-1em}
\end{figure}


Fig. \ref{PLoS curves} compares the $P_{LoS}$ against the 3D distance between ABS and GU in three standard urban environments using 200 randomly placed trees and 500 streetlights around the sidewalks. The green line indicates $P_{LoS}$ in the presence of buildings only, while red indicates additional $P_{LoS}$ by adding 200 trees in the city. The third blue line is $P_{LoS}$ in the presence of all obstacles. The results indicate that including 500 streetlights has a negligible effect on $P_{LoS}$, with an average difference of just 0.0170 between the red and blue curves. However, 200 trees have a significant impact, as reflected in Fig. \ref{Diff}. Therefore, the effect of streetlights is excluded in the remaining analysis. Similarly, the pink line compares $P_{LoS}$ findings in \cite{Saboor2023plos} for the ideal Manhattan environment with the random Manhattan environment presented in this paper. The results show that $P_{LoS}$ has identical trends in random and Manhattan environments for higher elevation angles.      

Fig. \ref{Diff} plots additional $P_{NLoS}$ from 200 trees in different urban environments. The results show that the influence of trees on $P_{LoS}$ is identical in urban and dense urban environments, primarily because the building's high difference is not significant in both environments. Also, we observe a max value of around 55$^\circ$ for both environments. During analysis, we analyze that for $\theta > 50^\circ$, mostly the last building between the ABS and the GU obstructs the LoS link, making $P_{LoS}$ more influenced by buildings at lower elevation angles ($\theta < 50^\circ$). In contrast, trees are more likely to obstruct LoS at higher elevation angles because they reside close to the GUs. However, as the elevation angle approaches 90$^\circ$, $P_{LoS}$ reaches nearly 100\% since both trees and buildings are less likely to interfere. In high-rise environments, the significantly taller buildings overshadow the influence of trees, reducing their impact on LoS obstruction until very high elevation angles. Fig. \ref{Density} illustrates $P_{LoS}$ against tree density, showing that tree density has a huge impact on $P_{LoS}$ in urban environments.   

\begin{figure}[!t]
  \centering
  \includegraphics[width=.75\linewidth]{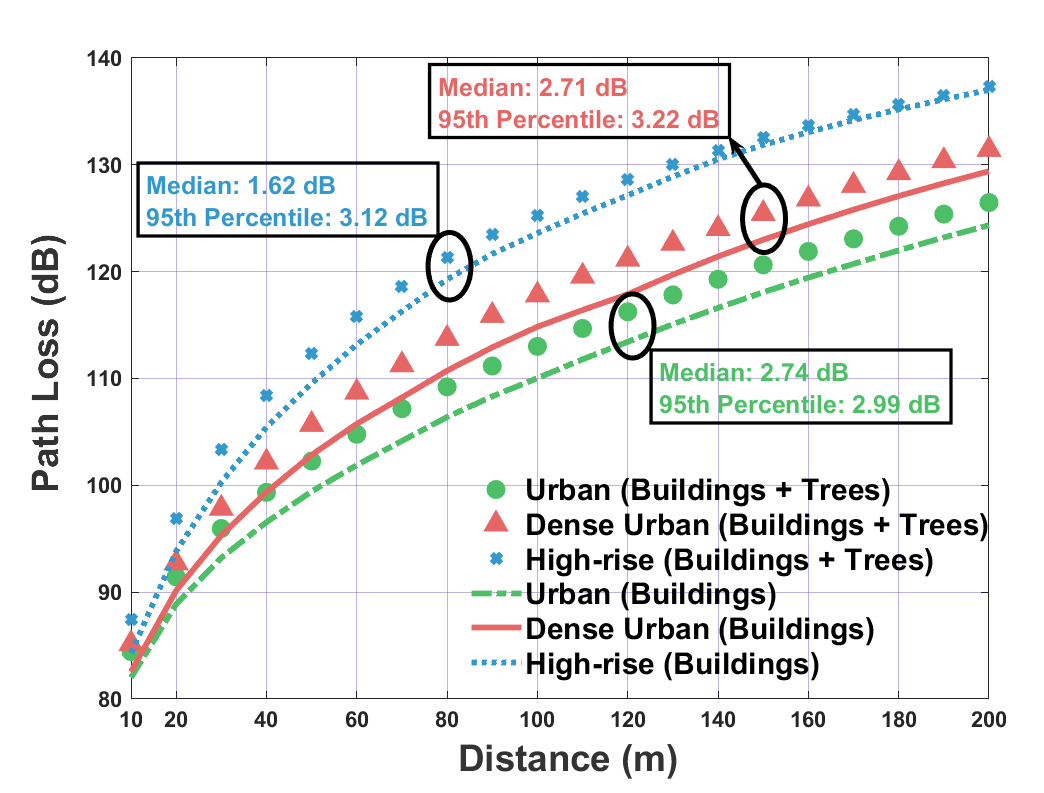}
  \caption{$PL$ in standard urban environments with 200 trees. }
  \label{PLd}
  \vspace{-1em}
\end{figure}

Fig.~\ref{PLd} compares the $PL$ (at 28 GHz) against varying 3D distances between ABS and GU in standard urban environments. Since trees are closer to users, we assume $d_2$ is randomly distributed between 4 and 8 meters and $d_1$ is the remaining 3D distance, i.e., $d_1 = d - d_2$. The markers represent $PL$ in the presence of buildings and trees, while the solid lines are $PL$ in the presence of buildings. Results show that 200 trees added a median $PL$ of 2.74~dB in urban, 2.71~dB in dense urban, and 1.62~dB in high-rise environments. This difference becomes more significant at higher elevation angles or shorter ABS distances to the GUs, where UAVs are more likely to operate. Also, it varies with more obstructions and higher frequencies. Thus, the impact of trees on $PL$ cannot be overlooked. 

\begin{figure}[!t]
  \centering
  \includegraphics[width=.75\linewidth]{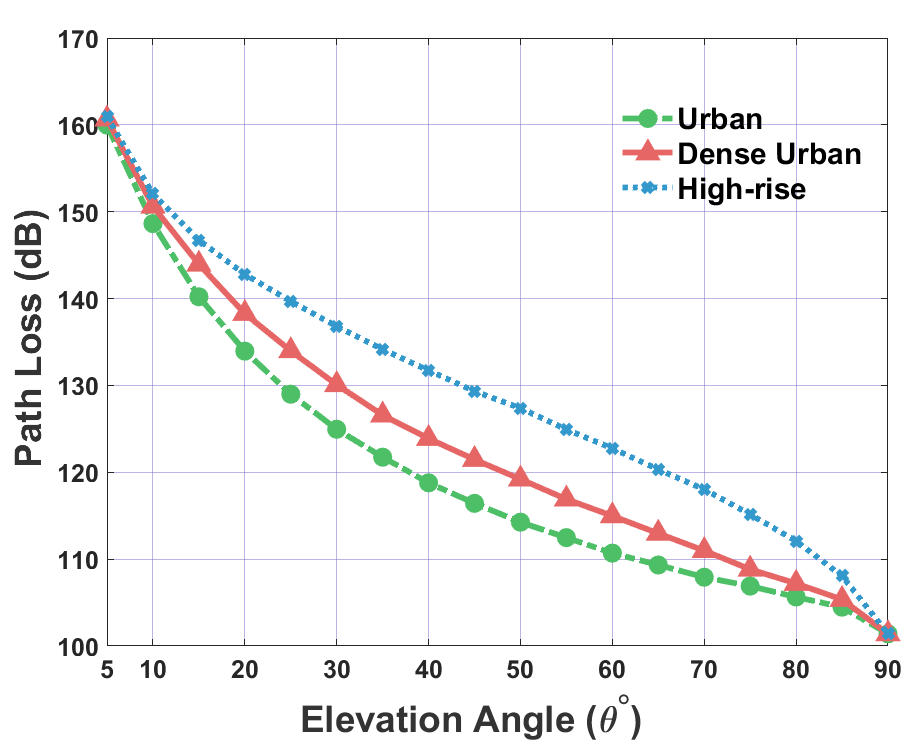}
  \caption{$PL$ against $\theta$ with $h_{ABS} = 100$~m. }
  \label{PLtheta}
   \vspace{-1em}
\end{figure}

For convenience, we derive the empirical $PL$ coefficient for equation \eqref{empirical} in Table \ref{tab2}, which estimates the $PL$ at 28~GHz in the presence of 200 trees across three standard urban environments. As visualized in the table, the path loss Exponent $B$ remains constant in an urban environment with or without trees, but $A$ changes. The primary reason is that the additional attenuation caused by trees is primarily distance-independent and can be treated as a constant offset to the total $PL$. The table also provides the Root Mean Square Error between the simulated results and the empirical model. Finally, Fig. \ref{PLtheta} further illustrates $PL$ as a function of $\theta$ in different urban environments, accounting for building and tree obstructions.


\section{Conclusion}
The paper presents a realistic MRS that evaluates $P_{LoS}$ in the general urban environments. In the proposed simulator, we first create a realistic urban layout by varying the shapes of buildings and incorporating trees and streetlights in the scenarios. Then, we estimate $P_{LoS}$ and $PL$ considering a complete path loss formulation. Results show that trees in the environment lead to lower $P_{LoS}$ and corresponding high $PL$. As an example, the median extra loss in the urban environment is 2.74~dB, with 2.99~dB in the 95th percentile in the urban environment. The proposed simulator, valuable insights, and empirical parameters will be useful in the design of UAV-based communication systems. 

\section*{Acknowledgment}
This research is supported by the Research Foundation
Flanders (FWO), project no. G0C0623N, and iSEE-6G project
under the Horizon Europe Research and Innovation program
with Grant Agreement No. 101139291.

\begin{table}[!t]
\caption{$PL$ fitting parameters for the standard urban environments at 28 GHz.}
\centering
\renewcommand{\arraystretch}{1.1} 
\begin{tabular}{|l|c|c|c|c|c|}
\hline 
\multirow{2}{*}{\textbf{Environment}} & \multirow{2}{*}{\textbf{B}} & \multicolumn{2}{c|}{\textbf{Buildings only}} & \multicolumn{2}{c|}{\textbf{Buildings + 200 Trees}}  \\ \cline{3-6} 
& & \textbf{A} & \textbf{RMSE} & \textbf{A} & \textbf{RMSE} \\ \hline
Urban & 3.38 & 43.90 & 1.84 & 46.55 & 1.60 \\ \hline
Dense Urban & 3.75 & 40.83 & 1.61 & 43.52 & 1.40 \\ \hline
High-rise & 4.26 & 38.64 & 0.99 & 40.26 & 1.15 \\ \hline
\end{tabular}
\label{tab2}
\vspace{-0.5cm}
\end{table}

\bibliographystyle{IEEEtran}
\bibliography{ref}

\end{document}